# Single-shot Compressed 3D Imaging by Exploiting Random Scattering and Astigmatism

Qiong Gao, * Weidong Qu, Ming Shao, Wei Liu, and Xiangzheng Cheng

*Luoyang Electronic Experiment Testing Centre, Luoyang, Henan 471003, China*
*Corresponding author: gaoqiong1980@126.com*

Based on point spread function (PSF) engineering and astigmatism due to a pair of cylindrical lenses, a novel compressed imaging mechanism is proposed to achieve single-shot incoherent 3D imaging. The speckle-like PSF of the imaging system is sensitive to axial shift, which makes it feasible to reconstruct a 3D image by solving an optimization problem with sparsity constraint. With the experimentally calibrated PSFs, the proposed method is demonstrated by a synthetic 3D point object and real 3D object, and the images in different axial slices can be reconstructed faithfully. Moreover, 3D multispectral compressed imaging is explored with the same system, and the result is rather satisfactory with a synthetic point object. Because of the inherent compatibility between the compression in spectral and axial dimensions, the proposed mechanism has the potential to be a unified framework for multi-dimensional compressed imaging.

Three-dimensional (3D) imaging based on compressed sampling (CS) theory is an active research area in recent years.[1-4] Compared with the traditional scanning/multi-shot and stereophotography methods,[5] CS method has the potential to enable the setup to be more compact. Compared with light field cameras,[6-7] the CS strategy can offer images of high resolution. Moreover, rather than the local refocusing implemented by light field cameras, depth sectioning can be achieved by exploring the sparse structure of the 3D field, which is desired in many applications. Motivated by the recent progress in imaging through scattering medium and point spread function (PSF) engineering, this Letter demonstrates a single-shot compressed incoherent 3D imaging mechanism.

Recent 3D imaging research benefits from the optical propagation through randomly scattering medium. Instead of regarding the scattering medium as an obstacle, the memory effect in random scattering can be utilized as an important tool to achieve imaging with diffraction-limited resolution and around corners.[8-9] The random medium can perform as an efficient analog multiplexer for light,[10] which is the core in designing compressed imaging system. With this desirable property, single-shot 3D[3,11-12] imaging has achieved by deliberately designed demultiplexing strategies and phase retrieval iterations. Another important strategy is to calibrate the imaging system by recording its PSFs at different axial locations modulated by scattering medium, and restores the 3D scene by solving an inverse problem.[4,13]

PSF engineering is a novel approach to achieve 3D super-resolution microscopy, which is realized by phase modulation to get a target PSF varying with axial location. The shapes of PSFs include ellipse,[14] double helix,[15] single-lobe,[16] saddle-point,[17] and self-bending.[18] In fact, this approach has also been explored in compressed imaging by producing a PSF with many random point responses.[19] In a more recent work, aberrations are utilized intentionally to generate PSFs that can simultaneously code and multiplex partial parts of the scene.[20]

Here we propose a single-shot compressed 3D imaging strategy which is implemented by combining 4f system and astigmatism together. A random phase mask is placed in Fourier plane of the 4f system, and the random modulation makes the PSF be a speckle pattern. The second lens of the 4f system is replaced by a pair of cylindrical lenses, and the induced astigmatism ensures the distinguishability among PSFs of different axial planes. Through an aforehand calibration of its PSFs, we can reconstruct 3D image from a single-shot compressed 2D raw image by solving an inverse problem with sparsity constraint.

The proposed method shows many distinct features or important extensions to previous works. In 3D super-resolution microscopy in Ref. 14, the astigmatism of a cylindrical lens is employed to obtain an elliptical PSF, whose peak position and peak width are used to determine the lateral and axial positions of a molecule. In our system, the astigmatism generated by a pair of cylindrical lenses is modulated further by a random phase, and the induced speckle-like PSF is more diverse in the sense of morphology, which makes it possible to measure the transverse and axial information simultaneously by deconvolution. The DiffuserCam in Ref. 4 relied on natural propagation of light to obtain different PSFs in axial dimension, which are enlarged or shrunk caustic patterns of a transparent phase mask. In contrary, the PSFs in our method enlarge in one direction and shrink in another direction simultaneously when scanning in a proper axial range, so the change of its shape is more evident, being a desired property in depth sectioning. In addition, the sensitivity of speckle to wavelength was utilized to design spectrometers of high-resolution.[21] Because the speckle was realized by optical propagation in a long fiber, its extension to imaging application is restricted.

The proposed optical configuration for single-shot compressed 3D imaging is shown in Fig. 1(a). The calibration of the imaging system is implemented by scanning a point source axially and recording the images. A LED white source is used to illuminate a pinhole with diameter of 10 μm, and the transmitted light is filtered and polarized sequentially. The band-passed and polarized light propagates through a collimated lens and a beam splitter, then arrives at a spatial light modulator (SLM, PLUTO-2-NIR-011, Holoeye Inc., 1920×1080 pixels). The pinhole and SLM are placed at the front and back focal planes of the lens, respectively. A random phase profile on SLM is generated by smoothing a 2D white noise. The modulated light is focused by a pair of cylindrical lenses and finally captured by a CMOS image sensor (IMX123LQT, SONY Inc., 2048×1536 pixels, sampling depth 12 bits). The two cylindrical lenses are placed closely, and play as the role of the second lens in traditional 4f system. As a preliminary step, an axial range for scanning is determined by addressing a plate phase profile on SLM. When the point source is moved gradually from the exact focal plane of the collimated lens, the spot captured by the image sensor changes approximately from a circle to an ellipse. The left and right scanning boundaries are reached when the elliptical spot degrades to a line. The right part of Fig. 1(a) shows such an evolution process when using a filter with central wavelength 530 nm (12 nm full width at half-maximum (FWHM)). An example of random phase used in our experiment is shown in the left part of Fig. 1(b), and its right column shows the modulated PSFs. The axial locations of PSFs in each column correspond to the unaffected ones in the right column of Fig. 1(a). It can be found that the random modulation in the Fourier plane scrambles the regular spot and produces many fine structures, which can be regarded as smoothed speckle patterns.

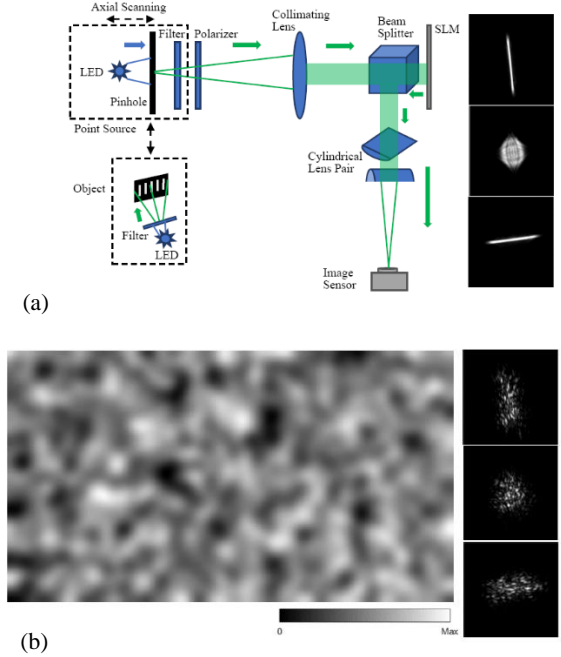

(a)

(b)

Fig. 1 (a) Optical configuration of the proposed single-shot compressed 3D imaging. The focal length of the collimating lens is 200 mm, and the distance between the two cylindrical lenses (also with focal length 200 mm in one direction) is 20 mm. The length of axial scanning scope is 20 mm with step 1mm. The right column shows three captured spots using a plate phase, from top to bottom, at left boundary, central point, and right boundary in axial direction. (b) An example of random phase (black to 0 and white to 1) addressed on SLM, and the induced PSFs corresponding to the axial locations in the right column of Fig. 1(a).

After the calibration procedure is completed, we replace the point source with a 3D object to carry out compressed imaging in a single-shot. The imaging process can be expressed mathematically by:

$$I_c(x,y) = \sum_{z_i} I_0(x,y;z_i) * h(x,y;z_i) \triangleq \mathbf{H} I_0 \quad (1)$$

where $I_0$ is the 3D object under acquisition, $h$ is PSF at each axial position, and $I_c$ is the captured 2D compressed image. Discrete variables are used in the expression, and the symbol * denotes convolution respect to $x/y$ variables. A convolution matrix $\mathbf{H}$ is introduced in Eq. (1) to express the process more compactly. The convolution model in Eq. (1) assumes implicitly the shift-invariance property of the imaging process. The field-of-view in our experiment is about 1.4°, limited by the size of CMOS pixel array. In addition, the small aperture is placed just before the SLM surface to fix a region of liquid pixels modulating the light. Therefore, the shift-invariance assumption is reasonably, and is also validated by moving the point source transversely.

When the calibrated PSFs and compressed image are available, the next core problem is how to reconstruct the 3D image $I_0$. We solve this ill-posed inverse problem by exploiting the hidden sparsity structure, and converting it to an optimization problem:

$$\hat{I} = \arg\min_{I \geq 0} \| I_c - HI \|_2^2 / 2 + \tau \| \Psi I \|_1 \quad (1)$$

where a transformation $\Psi$ is chosen to represent the image sparsely, $\tau$ is a regularization parameter, and the non-negativity constraint is forced for the solution. This problem has been investigated extensively in the applications of CS theory, and we use the alternating direction method of multiplier (ADMM).[22] Its application to the problem of form in Eqs. (1-2) is discussed exhaustively in Ref. 4 with $\Psi$ be 3D finite difference (FD) operator. Wavelet transform (WT) is used to represent the sparsity in this work, and 2D transform in $x/y$ plane is concerned in the implementation. Although larger computation burden, our experience indicates that the WT representation usually works better than the one with FD. The reconstruction procedure with WT is denoted as ADMM-WT solver, and the results in below are obtained with this solver after 1000 iterations and using wavelet of bior2.2 with level 4.

In preprocessing of the raw data, standard RGB data is obtained through format conversion and de-mosaicking. The data of G channel is selected and smoothed by a 4×4 average window, and then down-sampled by a factor 4. In addition, the background image is also collected, which is preprocessed similarly and then subtracted from the image containing object information. The PSF data is also normalized by its summation over all pixels. The lateral resolution degrades due to the pixel binning operation, but this will not affect the proof-of-concept demonstration of our imaging mechanism, which is the main purpose of this Letter. The depth of field is determined by the distance between the two cylindrical lenses, and the axial resolution is restricted by more factors, such as the just mentioned distance, the focal length of the cylindrical lens, and the scanning step in the calibration procedure. The above dependence offers more freedom to design a proper system for specific problem.

In order to reconstruct the image in each axial slice faithfully, the correlation between PSFs should be weak. To characterize this requirement quantitatively, we define a normalized correlation coefficient between 2 PSFs as

$$c_{z,z_0} = \max_{m,n}[h_z \star h_{z_0}](m,n) / \max_{m,n}[h_{z_0} \star h_{z_0}](m,n) \quad (3)$$

where the symbol $\star$ denotes cross correlation operation, the index $z$ denotes different axial location of PSF, and location $z_0$ is chosen as a reference. With the setup in Fig. 1, 21 PSFs with axial step 1 mm are collected, and the correlation coefficient curve is shown in Fig. 2(a). Although the curve is not strictly symmetric around the center, the coefficient decays quickly and drops to 0.6 approximately at the axial boundaries.

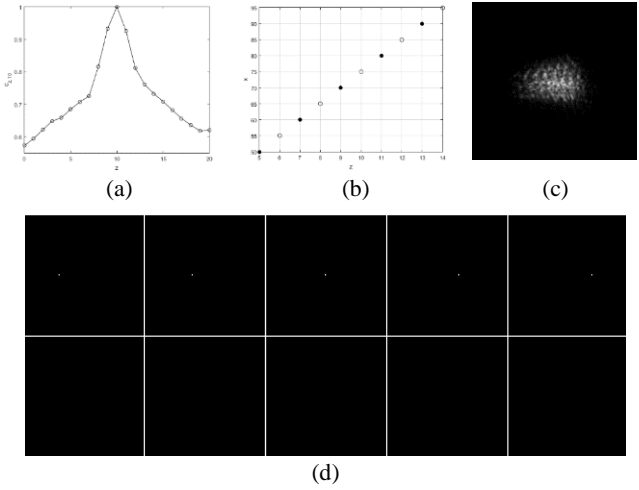

Fig. 2 (a) Correlation coefficients of PSFs in 21 axial planes with step 1 mm.
(b) Construction of a 3D point object with solid circles.
(c) Simulated compressed image of a synthetic 3D point object with 5 points corresponding to the solid circles in Fig. 2(b).
(d) Reconstructed images in 10 axial planes with ADMM-WT solver (top to bottom and left to right.). The object in the top row only occupies one pixel in each subfigure.

As a preliminary but important demonstration of the proposed imaging strategy, a synthetic 3D point object is constructed, which consists of 5 points shown in Fig. 2(b) by solid circles in $xz$ space with step of 10 pixels in $x$ direction. With 10 PSFs ($z = 5$~$14$) to decode this compressed image, ADMM-WT solver gives the reconstructed image in each axial plane as shown in Fig. 2(c). For the experiment and the solver, it is a much more rigorous validation to include neighboring PSFs to reconstruct the object. If the axial sectioning capability is not enough, artificial objects may appear in the neighboring planes. The obtained image is rescaled linearly to [0, 1] within the whole 3D data cube. Fig. 2(c) shows five separate points in the top row with peak pixel values 0.99/0.99/1.00/0.97/0.97, and black background in the bottom row with peak pixel values of order $10^{-2}$ or even smaller. So the five points in the relevant axial planes are restored accurately, and the information linkage to neighboring axial planes is negligible.

Now we demonstrate the strategy with real experiment. The marks on a tilted ruler shown in Fig. 3(a) is taken as a 3D object when the axis of the imaging system is not orthogonal to its surface. The ruler has scales of 1 mm (bottom part) and 0.9 mm (top part), and the angle between its normal direction and the axis of the imaging system is about 75°. With the setup and phase mask in Fig. 1, we obtain the compressed image shown in Fig. 3(b). In order to obtain reconstructed image of high quality, it is crucial to employ a proper subset of PSFs for the solver. After some trials, a group of PSFs with $z = 8$~$15$ is utilized to decode the compressed image. The obtained results are shown in Fig. 3(c-d). The image in Fig. 3(c) is synthesized by adding the reconstructed 3D image along axial direction, which offers a global impression of the scene. The eight 2D images in the 3D data cube are shown separately in Fig. 3(d) to highlight the depth sectioning capability of the proposed imaging strategy. Comparing the subfigures in Fig. 3(d) and Fig. 3(c), we can find that the lines in different axial planes appear in sequential slices, as if obtained by illuminating each axial planes separately.

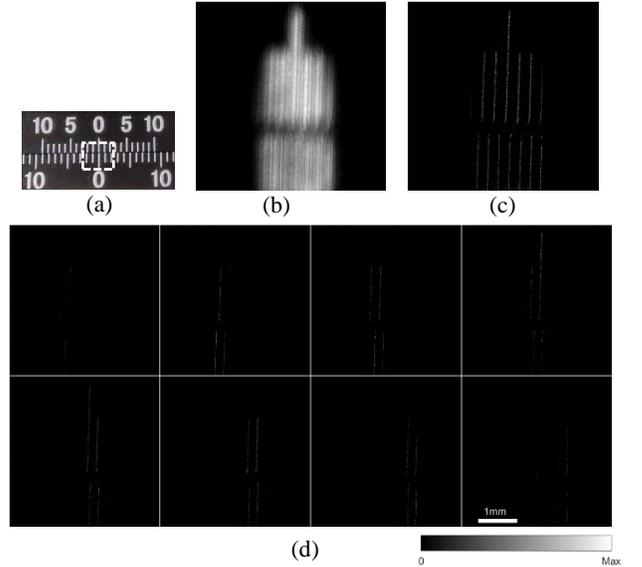

Fig. 3 Experimental demonstration of the proposed 3D compressed imaging mechanism.
(a) Part of a ruler used in the experiment. The dashed box shows the illuminated area. The upper part is shifted about 0.5 mm and the image is inverted in experiment.
(b) Compressed image obtained with the setup in Fig. 1.
(c) Reconstructed 2D image obtained by projecting the 3D image along axial direction.
(d) Eight axial slices of the reconstructed 3D image (left to right and top to bottom).

Finally, we explore the potential application of our imaging mechanism to 3D multispectral imaging. Using spectral filter in Fig. 1(a) with central wavelength 510 nm (FWHM = 8 nm) and 578 (FWHM = 9 nm) nm, respectively, the calibration procedure is repeated and two more groups of PSFs are obtained. Using similar method to Fig. 2(b), we can construct a 3D multispectral point object with these PSFs at different wavelengths but registered spatially. Six PSFs denoted by solid circles in Fig. 4(a) are utilized to synthesize such an object, and the simulated compressed image is shown in Fig. 4(b). To encode this compressed image with spectral information, the imaging model in Eq. (1) has to be generalized so that the summation is over relevant axial-spectral channels. With the 12 PSFs denoted in Fig. 4(a), ADMM-WT solver gives the results in Fig. 4(c), which displays the reconstructed 2D images in each axial-spectral channel (rescaled linearly to [0, 1] in the whole data matrix). The peak pixel values for 578 nm are 0.95/0.90/0.04/0.01 from left to right, 0.05/0.77/0.69/0.09 for 530 nm, and 0.07/0.08/0.88/1.00 for 510 nm. It can be found that the point in each channel is restored accurately and the SNRs are also quite satisfactory. Although this object is simple and sparse (including the one in Fig. 2), the above results imply potential application of the proposed system to military surveillance on long-distance threats when combined with a telescope. Compared with other compressed spectral imaging methods based on coded aperture,[23] spectral filter array,[24] or speckles,[25] our method just employs the natural disperse characteristics of common lenses and SLM. Rather than combining different compression mechanism in different dimensions directly,[26] the spectral compression is compatible inherently with the axial compression in our system.

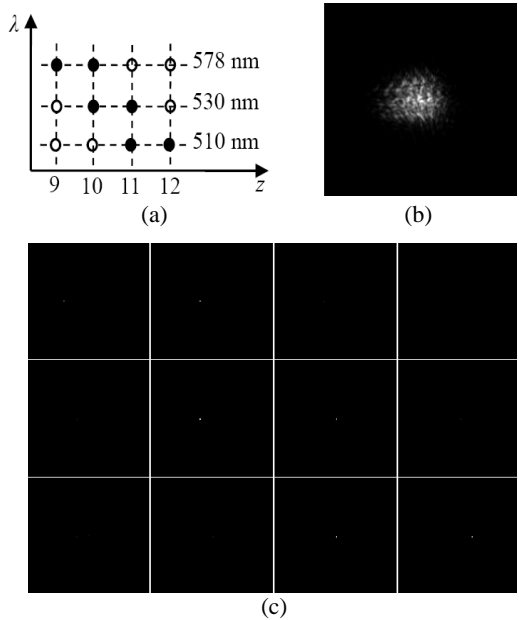

Figure 4 Demonstration of the proposed imaging mechanism with a synthetic multispectral 3D point object.
(a) Construction of a Part of a multispectral 3D point object. The black circles denote the used PSFs.
(b) Simulated compressed image with the six PSFs in Fig. 4(a).
(c) Reconstructed 2D images in different axial-spectral channels. The object in each relevant channel concentrates mainly on one pixel in each subfigure.

In conclusion, a novel single-shot compressed 3D imaging strategy is proposed, which leverages the random phase modulation in Fourier plane and the astigmatism induced by a pair of cylindrical lenses. The design leads to speckle-like PSFs sensitive to axial locations, which makes it be feasible to reconstruct the 3D image by solving an inverse problem. With the calibrated PSFs, the imaging strategy is demonstrated with a synthetic 3D point object and a real 3D object, and the ADMM-WT solver can reconstruct the images in different axial slices faithfully. Moreover, the application of our compression mechanism to 3D multispectral imaging is further explored with a synthetic object, and the reconstructed results are encouraging. Although the present Letter only concerns 3D/multispectral imaging, it is possible to achieve ultrafast compressed imaging by introducing a galvanometer scanner into the setup. Therefore, the compressed imaging mechanism based on random scattering and astigmatism is a promising candidate for realizing single-shot multi-dimensional imaging in some combination domains among space, time, and spectrum.